\documentclass[aps,prl,twocolumn,preprintnumbers,superscriptaddress,floatfix,nofootinbib,notitlepage,showkeys,showpacs]{revtex4-1}

\usepackage[utf8x]{inputenc}
\usepackage{titlesec}
\usepackage{graphicx}
\usepackage{hyperref}
\usepackage{latexsym}
\usepackage{amsmath}
\usepackage{amssymb}
\usepackage{bbm}
\usepackage{ulem}
\usepackage{pdfsync}
\usepackage{epsfig}
\usepackage{epstopdf}
\usepackage{color}
\usepackage{comment}
\usepackage{placeins}
\usepackage{slashed}
\usepackage{mathrsfs}

\newcommand{\LG}{L}
\newcommand{\DD}{D}

\definecolor{lmpurple}{RGB}{151, 9, 222}

\def\lsim{\mathrel{\raise.3ex\hbox{$<$\kern-.75em\lower1ex\hbox{$\sim$}}}}
\def\gsim{\mathrel{\raise.3ex\hbox{$>$\kern-.75em\lower1ex\hbox{$\sim$}}}}

\titleformat{\section}[runin]{\normalfont\scshape}{\thesection}{1em}{\textbf}

\begin{document}

\title{Search for Dark Matter Effects on Gravitational Signals from Neutron Star Mergers}

\author{John Ellis}
\email{john.ellis@cern.ch}
 \affiliation{NICPB, R\"avala pst.~10, 10143 Tallinn, Estonia}
\affiliation{ Theoretical Particle Physics and Cosmology Group, Physics Department, King's College London, London WC2R 2LS, United Kingdom}
 \affiliation{Theoretical Physics Department, CERN, Switzerland}
 \author{Andi Hektor}
 \email{andi.hektor@cern.ch}
 \affiliation{NICPB, R\"avala pst.~10, 10143 Tallinn, Estonia}
 \author{Gert H{\"u}tsi}
 \email{gert.hutsi@to.ee}
 \affiliation{NICPB, R\"avala pst.~10, 10143 Tallinn, Estonia}
 \affiliation{Tartu Observatory, T{\~o}ravere 61602, Estonia}
 \author{Kristjan Kannike}
 \email{kristjan.kannike@cern.ch}
 \affiliation{NICPB, R\"avala pst.~10, 10143 Tallinn, Estonia}
 \author{Luca Marzola}
\email{luca.marzola@cern.ch}
 \affiliation{NICPB, R\"avala pst.~10, 10143 Tallinn, Estonia}
  \author{Martti Raidal}
  \email{martti.raidal@cern.ch}
 \affiliation{NICPB, R\"avala pst.~10, 10143 Tallinn, Estonia}
 \author{Ville Vaskonen}
\email{ville.vaskonen@kbfi.ee}
 \affiliation{NICPB, R\"avala pst.~10, 10143 Tallinn, Estonia}

\begin{abstract}
{Motivated by the recent detection of the gravitational wave signal emitted by a binary neutron star merger, 
we analyse the possible impact of dark matter on such signals. We show that dark matter cores in merging 
neutron stars may yield an observable supplementary peak in the gravitational wave power spectral density
following the merger, which could be distinguished from the features produced by the neutron components.}\\

\end{abstract}
\begin{centering}
\leftline{CERN-TH-2017-208, KCL-PH-TH/2017-50}
\end{centering}

\maketitle

\section{Introduction.}

The discovery of gravitational waves (GW) by the LIGO and VIRGO collaborations has opened a new 
observational window on the Universe. The first measured GW signals were emitted from the merging of 
black hole binaries~\cite{Abbott:2016blz,Abbott:2016nmj,Abbott:2017oio} and spectacularly confirmed the predictions of general relativity and constrained its extensions in the strong-field 
regime~\cite{TheLIGOScientific:2016src,Yunes:2016jcc}. Recently the LIGO and VIRGO collaborations have also 
detected a GW signal originating during the inspiral leading to the merger of two neutron stars (NS), GW170817~\cite{Ligo+Virgo}.
On this occasion, the observation of the GW 
signal was accompanied by counterparts in many bands of the electromagnetic spectrum~\cite{Everybody}, 
as was expected for a NS-NS merger, notably including the observation of
a gamma-ray burst, GRB170817A~\cite{LIGO+Fermi+Integral}. 
This new discovery is an important confirmation that
NS-NS mergers are the originators of short gamma-ray bursts~\cite{1986ApJ...308L..43P,1989Natur.340..126E,1992ApJ...395L..83N,Berger:2013jza}, 
and play a major role in the r-process nucleosynthesis as demonstrated by the ejection of synthesised material via subsequent kilonova explosions \cite{Li:1998bw,Goriely:2011vg,Tanvir:2013pia,Metzger:2014yda,2016Natur.531..610J,Metzger:2016pju}.

Conventional models of NS-NS merger yields three separate peaks
in the GW emission following the merger~\cite{Oechslin:2007gn,Bauswein:2011tp,Bauswein:2012ya,Takami:2014zpa,Takami:2014tva,Foucart:2015gaa,Bauswein:2015yca,Clark:2015zxa,Radice:2016rys,Bose:2017jvk}: for a review with a comprehensive list of references, see~\cite{Baiotti:2016qnr}. The strongest observable peak
in these simulations results from the rotation of the bar-deformed hyper-massive neutron star 
in the first instants after the merger. This strong peak is flanked by two less pronounced peaks
that are produced by the dynamics of the remnants of the original stellar cores. Despite
the complexity of the simulations, the above-mentioned qualitative features of the GW signal 
are captured by a simple mechanical model \cite{Takami:2014zpa,Takami:2014tva} that reproduces
qualitatively the detailed calculation of the GW waveform emitted in these events. 
In the case of GW170817, no post-merger GW signal was observed, but future
observations of such signals would have the capability to constrain models of dark matter
as well nuclear matter, as we now discuss.

We show in this Letter how observations of GW signals from NS-NS mergers also offer a
new window on the dark matter (DM) component of the Universe~\cite{Ade:2015xua}, through
a possible supplementary peak in the power spectral density (PSD) of the GW emission
following a NS-NS merger. We introduce additional features into the simple mechanical model of NS-NS mergers
to account for the possible presence of DM cores within the neutron stars. We show that these
DM cores may generate an additional well-defined peak in the PSD of the GW signal emitted after
the merger. If the DM cores contribute about $5 - 10\%$ of the NS mass, we find that the DM peak is observable and can be clearly distinguished from the features of the signal induced by the remnant neutron matter cores for a wide range of effective parameters. 

It should be noted that ordinary WIMP DM cannot condense inside stars or compact stellar remnants 
in quantities as large as discussed in this study. For example, during a period of $\sim 10$\,Gyr, 
even after allowing for DM densities $\sim 10^4$ times larger than the local value 
(as may be reachable in the central parts of the Galaxy), the typical amount of accreted WIMP DM 
does not significantly exceed $\sim 10^{-10} M_{\odot}$~\cite{Goldman:1989nd,Kouvaris:2007ay,Kouvaris:2010vv,Guver:2012ba}. 
However, having a nonstandard dark sector with dissipative (but  subdominant) DM component 
(e.g., as in~\cite{Foot:2004pa,Fan:2013yva,Pollack:2014rja}) might lead to the formation of DM-admixed 
stars with noticeable DM content. For example, DM-admixed NSs have been discussed previously
in~\cite{Sandin:2008db,Leung:2011zz,Li:2012qf,Goldman:2013qla,Leung:2013pra,Mukhopadhyay:2016dsg,Panotopoulos:2017idn}. 
Moreover, the formation of asymmetric DM stars, which might potentially serve as cores for the NSs studied in this paper, 
have been investigated in~\cite{Kouvaris:2015rea,Maselli:2017vfi}. Although the details how these objects might have formed 
require much deeper investigation, which is beyond the scope of this study, we propose the following illustrative scenario. 
If a small fraction of DM ($\lesssim 10\%$) is strongly self-interacting then, as demonstrated in~\cite{Pollack:2014rja}, 
it can lead to early formation of black holes (BHs) that can serve as seeds for subsequent supermassive BH formation,
alleviating tension with the small amount of time available for their formation in the standard LCDM picture. 
Depending on the shape of the initial perturbation spectrum, in addition to BH formation, 
one would also expect formation of smaller structures analogous to visible-sector NSs. 
Later on, these compact dark objects can serve as accretion centers for baryonic matter, 
which when accreted in sufficient quantities can lead to the formation of DM admixed NSs. 
For this scenario to work no interaction beyond gravity is needed between the dark and the visible sectors.

The exploratory results in this Letter clearly indicate that observations of NS-NS mergers 
have the potential to probe the properties of such self-interacting DM, complementing dedicated searches. 
Our work underlines the need for further simulations to assess with more precision the 
possible impact of a DM component on the evolution of NS-NS mergers and disentangle
its effects from phenomena associated with different nuclear equations of state.

\section{A mechanical model.}
\begin{figure}
\centering
\includegraphics[width=.6\linewidth]{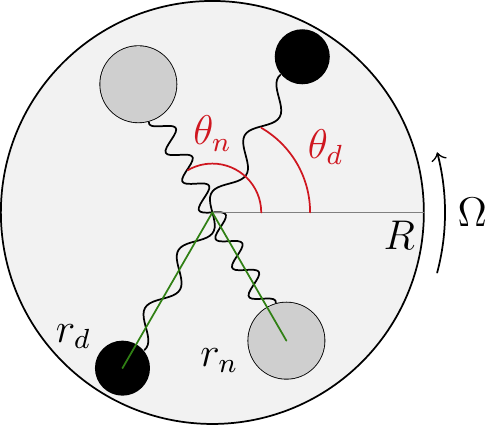}
\caption{A schematic diagram of the mechanical model adopted to investigate the possible
effects of DM components on the GW signal emitted by NS-NS mergers. The DM cores (in black) 
move in an environment constituted of the remnant neutron stellar core (in white) 
and a disk of neutron matter enclosing the latter.}
\label{fig:KKart}
\end{figure}

In order to assess the impact of a possible DM component on the GW
signal emitted by a NS-NS merger, we extend the mechanical model originally presented 
in~\cite{Takami:2014zpa,Takami:2014tva}. We include two additional DM cores of 
(for simplicity) equal mass $m_d/2$, which move in the gravitational potential of the hyper-massive 
NS generated in the merger. The latter is modelled as a disk of mass $M$ containing
two coupled oscillators of (again for simplicity) equal mass $m_n/2$, which represent the neutron 
components of the remnant stellar cores, as illustrated in Fig.~\ref{fig:KKart}. 
The disk and the neutron stellar cores rotate at a common angular velocity $\Omega=\dot\theta_n$, 
whereas the dark cores rotate at an angular velocity $\dot\theta_d$. We account for differences in the strength of DM and neutron interactions by considering different spring constants and damping factors. 

The Lagrangian of the mechanical system takes the form
\begin{equation}
\begin{aligned}
\LG = \frac{m_n}{2} \left(\dot r_n^2+\left(r_n \dot\theta_n\right)^2\right) + \frac{m_d}{2} \left(\dot r_d^2+\left(r_d \dot\theta_d\right)^2\right) \\+ \frac{M R^2 \dot\theta_n^2}{4} + 2 k_n (r_n-a_n)^2+2 k_d (r_d-a_d)^2 \,,
\label{L}
\end{aligned}
\end{equation}
where an overdot indicates differentiation with respect to time and the subscripts $n$ and $d$
distinguish quantities related to the nuclear and dark cores, respectively. The parameters
$k_n$ and $k_d$ are effective `spring constants' that characterize the oscillations of
the nuclear and DM cores in the gravitational potential and the medium. The effective damping 
induced by interactions is encoded in the parameters $b_n$ and $b_d$, and
that due to GW emission, $c_n$ and $c_d$, is included by adding the term
\begin{equation}
\begin{aligned}
\DD =& -b_n \dot r_n^2 - b_d \left(\dot r_d^2+\left(r_d \left(\dot\theta_n - \dot\theta_d\right)\right)^2\right) \\ &- c_n \left(\dot r_n^2+\left(r_n \dot\theta_n\right)^2\right) - c_d \left(\dot r_d^2+\left(r_d \dot\theta_d\right)^2\right) \,.
\end{aligned}
\end{equation}
The equations of motion for the relevant degrees of freedom are obtained
from the dissipative Euler-Lagrange equation
\begin{equation}
\frac{{\rm d}}{{\rm d}t} \frac{\partial \LG}{\partial \dot q_j} - \frac{\partial \LG}{\partial q_j} = \frac{\partial \DD}{\partial \dot q_j} \,.
\end{equation}

In order to investigate the resulting GW signal, we compute the strain induced by the 
plus and cross polarisations at a distance $D$ from the merger  
\begin{equation}
h_+ = \frac{\ddot I_{xx} + \ddot I_{yy}}{D} \,,\quad h_\times = \frac{\ddot I_{xy}}{D} \,,
\end{equation}
where $I_{kl}$, $k,l\in\{x,y,z\}$, are the components of the quadrupole-moment tensor:
\begin{equation}
\begin{aligned}
&I_{kl} = \sum_{j=n,d} \left(\bar I_{j,kl} - \frac{1}{3}\delta_{kl} \delta^{ab} \bar I_{j,ab}\right) \,, \\ &\bar I_{j,kl} = 3m_j x_{j,k} x_{j,l} \,.
\end{aligned}
\end{equation}
The PSD is then obtained as
\begin{equation}
\tilde h(f) = \sqrt{\frac{|\tilde h_+(f)|^2 + |\tilde h_\times(f)|^2}{2}} \,,
\end{equation}
where the tilded quantities are the Fourier transforms of the corresponding GW strains.

\section{The effect of dark matter.}

\begin{figure}
\centering
\includegraphics[width=.9\linewidth]{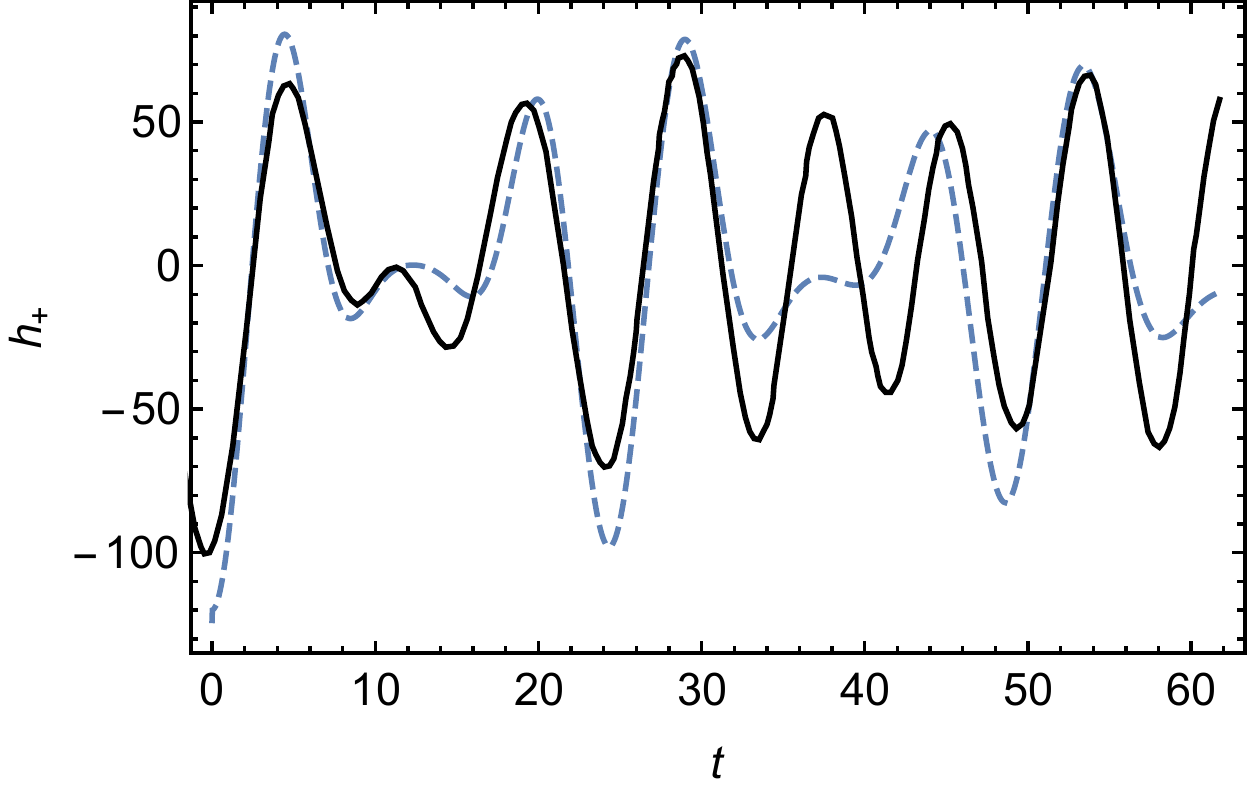}
\caption{Comparison between the results for the strain $h_+$ calculated in the
full numerical-relativity simulation reported in ~\cite{Takami:2014zpa,Takami:2014tva} (black solid line)
with results from the simple mechanical model in (\ref{L}) (blue dashed line).}
\label{fig:comparison}
\end{figure}

As was shown in~\cite{Takami:2014zpa,Takami:2014tva}, the simple mechanical model without the
DM components reproduces semi-quantitatively the peaks in the PSD found in a detailed simulation including
general-relativistic and nuclear effects. Fig.~\ref{fig:comparison} shows that, for suitable values
of the model parameters, our implementation
also reproduces well the full numerical-relativity strain $h_+$ for the full simulation reported in
~\cite{Takami:2014zpa,Takami:2014tva}. This comparison gives us confidence that such a simple
model can capture correctly the dominant dynamical effects.

\begin{figure}
\includegraphics[width=0.94\linewidth]{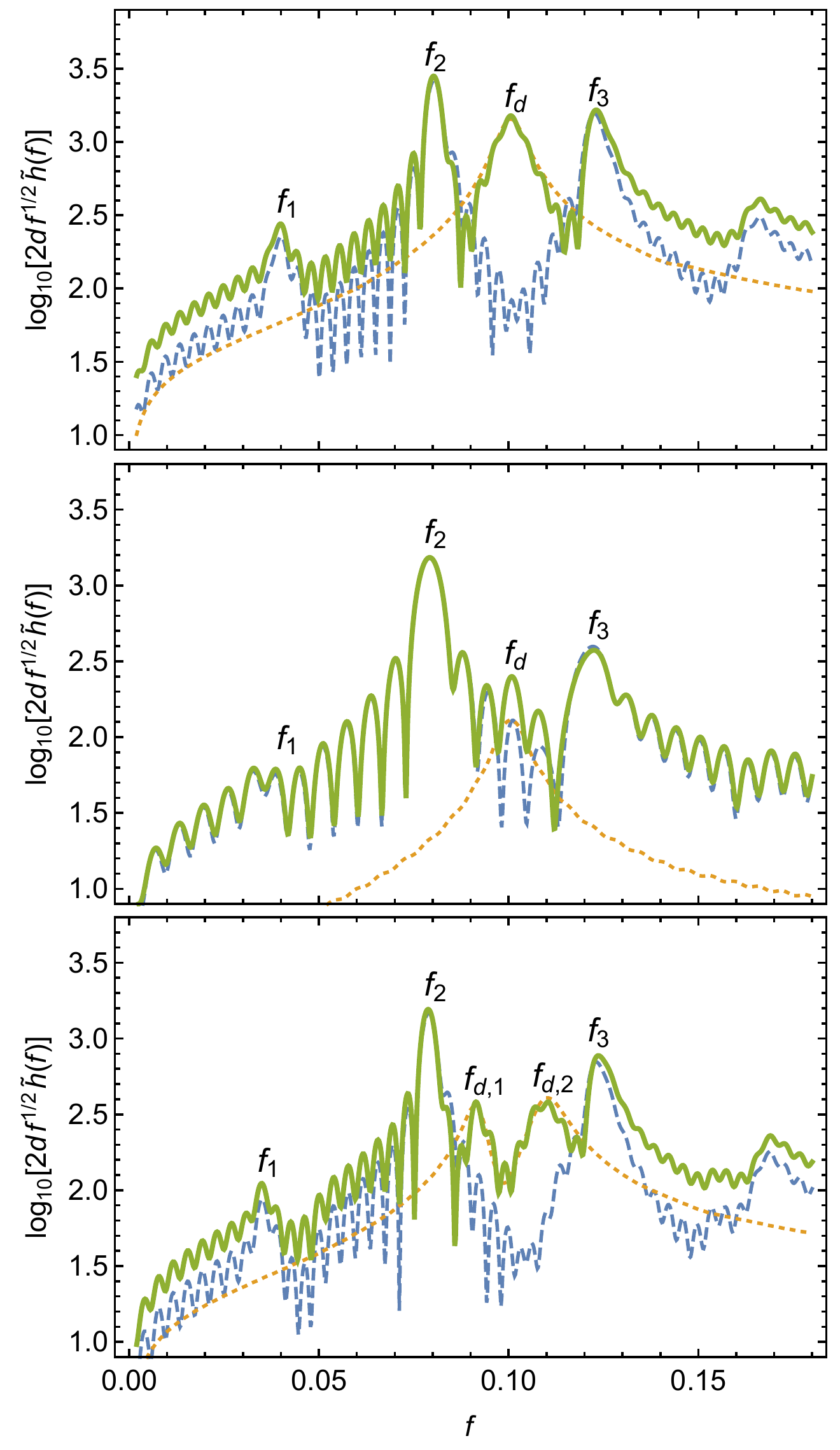}
\caption{Top panel: the gravitational wave (GW) power spectral density (PSD)
immediately following the merger of two equal-mass neutron stars. Middle panel: the GW PSD of the same merger
during a later period. Bottom panel: the GW PSD immediately following the merger of two neutron stars
of unequal masses.
The solid green lines show the full GW power spectral density, whereas the dashed blue and 
dotted yellow lines represent the contributions of the nuclear matter and DM, respectively. The units are arbitrary.
Note the splitting of the DM contribution in the unequal-mass case.}
\label{fig:GW}
\end{figure}

The PSD of the GW emission resulting from the above model (\ref{L}) including the DM
components is shown in Fig.~\ref{fig:GW}. The top panel shows the PSD immediately
following the merger of two equal-mass neutron stars, whereas the middle shows the PSD during a later time period.
In calculating this representative point we set the 
parameters of the nuclear matter to $M=10$, $m_n=10$, $k_n=0.25$, $b_n=0.1$, $c_n=0.02$ and $a_n=1$,
and the DM ones to $m_d=2$, $k_d=0.05$, $b_d=0.02$, $c_d=0.02$ and $a_d=0$.
The assumption that $a_n \ne 0$ was motivated by the presence of a hard repulsive nuclear core,
whereas we expect no repulsive DM interactions and set $a_d = 0$. As initial conditions we adopted 
$\dot r_n(0) = \dot r_d(0) = 0$, $\dot \theta_n(0) = \dot \theta_d(0) = 0.2$, $r_n(0)=R/2$ and $r_d(0)=R$, 
and the equations of motion were solved on a time interval $t\in[0,260]$ in arbitrary units. 

The blue dashed line in the top panel of Fig.~\ref{fig:GW} shows the PSD of the GW signal generated by the
nuclear matter immediately following the equal-mass NS-NS merger. Three peaks are clearly visible: the lowest frequency
peak, at $f = f_1$, correlates well with the compactness of the final object in a way 
that is insensitive to the neutron equation of state. The tallest peak, at $f = f_2$,
is induced by the rotation of the hyper-massive neutron star and is sensitive to the equation of state. 
As DM cores and associated DM-neutron interactions could induce effective
modifications of the latter, we expect that measurements of the neutron matter PSD alone could 
already provide important information on the impact of DM on the merger dynamics, if the nuclear
uncertainties in the equation of state are under control. The third peak, at $f=f_3$,
is generated together with the $f_1$ peak in the dynamics of the remnant NS cores,
and generally falls outside the current observational window~\cite{Takami:2014zpa,Takami:2014tva}. 

The impact of DM is signalled in the top panel of Fig.~\ref{fig:GW} by the yellow dotted line, 
which shows the peak generated by this component at a frequency $f_d\propto \sqrt{k_d/m_d}$. 
The continuous green line simply illustrates the sum of the two contributions, with the DM
contribution being clearly visible in this particular example, though we caution that its
position and height are dependent on unknown aspects of the DM dynamics.

The middle panel of Fig.~\ref{fig:GW} shows the PSD of the same equal-mass GW signal for a later period ($t\in[100,260]$)
of emission. We see that the $f_1$ peak has reduced, an effect seen already in Fig.~9 of~\cite{Takami:2014tva},
as has the DM peak. This is because the DM cores relax relatively rapidly, in particular via
GW emission, whereas the relaxation of the neutrons component is resisted by the remnant 
repulsive neutron cores.

The bottom panel of Fig.~\ref{fig:GW} shows the PSD of the GW signal from the merger of a pair of
neutron stars with unequal masses ($m_1/m_2 = 0.7$). In this case we see that the peak of the DM contribution is split,
whereas the peaks due to nuclear matter are not split. This splitting of the DM signal would be an
interesting diagnostic tool in the event that a DM signal is detected.

\begin{figure}
\vspace{0.5cm}
\includegraphics[width=0.9\linewidth]{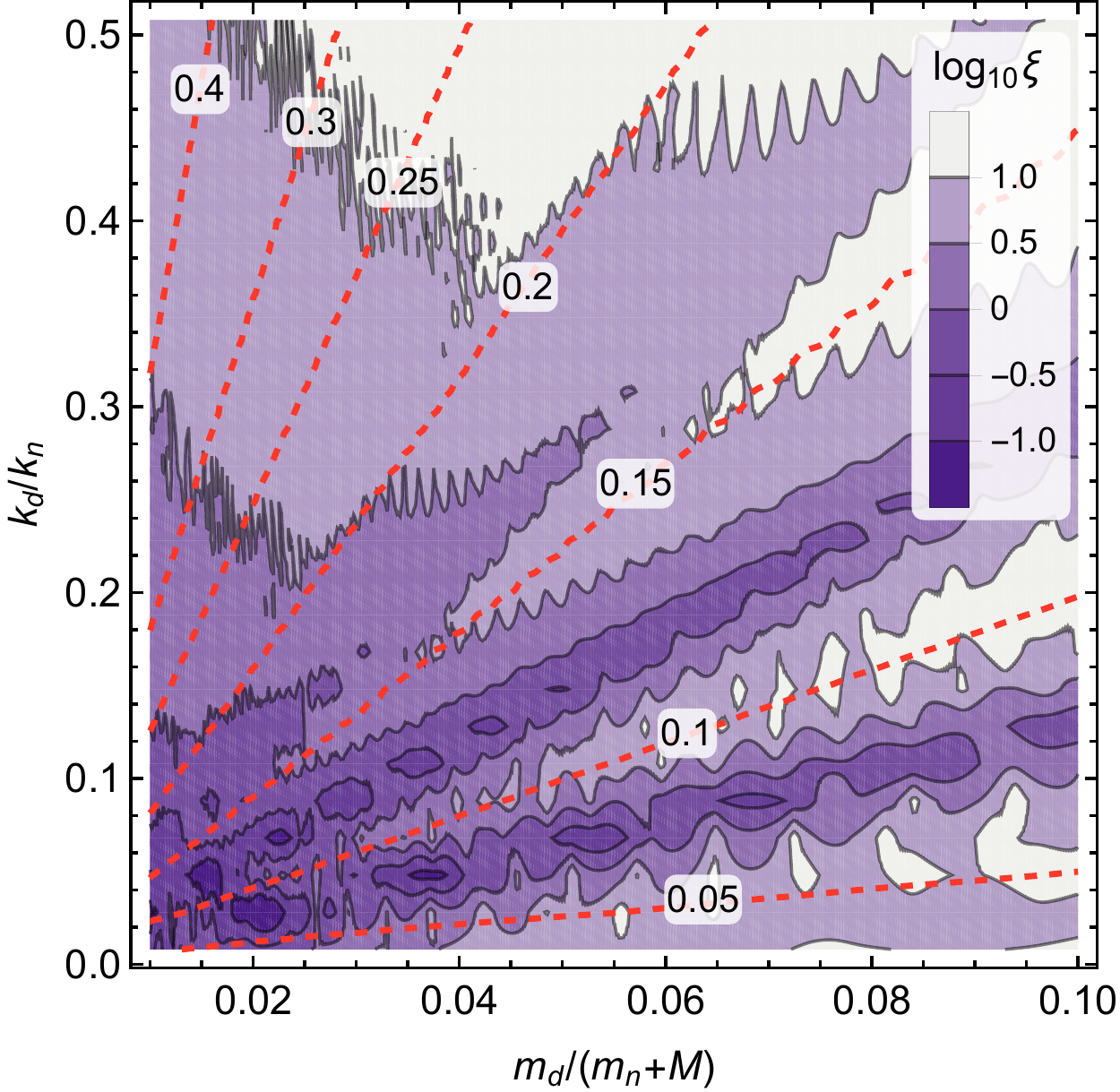}
\caption{The relative strength $\xi$ (\protect\ref{xi}) of the DM component to the GW strain amplitude
is indicated by colour coding. The dashed red contours show the peak frequency of the DM contribution 
in the same arbitrary units as in Fig.~\ref{fig:GW}.}
\label{fig:xi}
\end{figure}

The dependence of the DM signal strength on the effective parameters is encoded in the quantity 
\begin{equation}
\xi = \left(\frac{r_d(0)}{R}\right)^2 \frac{\tilde h_d(f_d)}{\tilde h_n(f_d)} \,,
\label{xi}
\end{equation}
which describes the strength of the DM contribution to the GW strain, at the peak frequency $f_d$, 
relative to the contribution of the nuclear matter. Fig.~\ref{fig:xi} illustrates the dependence of the 
$\xi$ parameter on the rescaled effective parameter adopted, demonstrating that the DM peak 
potentially leaves a distinguishable signature in the GW PSD over large ranges of parameter values. 
In making this plot we used, apart from $m_d$ and $k_d$, the same parameters and initial conditions 
as in the top panel of Fig.~\ref{fig:GW}. The red dashed lines illustrate the peak frequency contours of the DM contribution.   

\section{Discussion and conclusions.}
\label{sec:discussion}

We have considered the possible effect of a dark matter component on the gravitational wave 
signal emitted in a neutron star-neutron star merger event. We have extended for this purpose
a simple mechanical model proposed in~\cite{Takami:2014zpa,Takami:2014tva} that captures
essential features of the dynamics of the hyper-massive neutron star formed in the first instant 
after the merger. For the purpose of our analysis we allowed significant fractions of the 
original neutron star masses -- up to about $10\%$ -- to be in the form of dark matter cores. 

According to our results, the dark matter cores may produce a supplementary peak in the characteristic 
gravitational wave spectrum of neutron star mergers, which can be clearly distinguished from the features 
induced by the neutron components. Whilst precise simulations of merging neutron stars are 
certainly needed to fully quantify the effect, the emergence of the new peak demonstrates that 
analyses of these gravitational wave signals have the potential to shed new light on the properties of dark matter. Depending on the formation history, the sizes of the dark cores may vary considerably, thus the location and the amplitude of the new peak is expected to change. This non-universal behavior potentially helps distinguish the new dark matter peak from the others produced by baryonic physics. Thus, future observations of the GW signals from NS-NS mergers could provide interesting insight into dark matter physics, as well as into gravitation, nuclear physics and astrophysics.

\section{Acknowledgements.}
We thank K. Takami, L. Rezzolla, and L. Baiotti for useful discussions. This work is supported by the Estonian Research Council grant MOBTT5, ST/L000326/1, IUT23-6, IUT26-2, PUT1026, PUT799 and ERDF Centre of Excellence project TK133. 
\vspace{-.8cm}
\bibliography{peaks}
\end{document}